\documentclass{iau}
\usepackage{graphicx}
\usepackage{natbib}
\usepackage{aas_macros}
\usepackage{amsmath}
\title[Predicting the Emergence of Solar Active Regions Using Machine Learning] 
{Predicting the Emergence of Solar Active Regions Using Machine Learning}

\author[S. Kasapis et al.]   
{Spiridon Kasapis$^1$, Irina N. Kitiashvili$^1$, Alexander G. Kosovichev$^{1,2}$, John T. Stefan$^2$ \and Bhairavi Apte$^2$}

\affiliation{$^1$NASA Advanced Supercomputing Division, NASA Ames Research Center, \\
N258, Moffett Field, CA 94035, United States \\[\affilskip]
$^2$Department of Physics, New Jersey Institute of Technology, \\ University Heights, Newark, NJ 07102, United States}

\pubyear{2024}
\volume{365}  
\pagerange{119--126}
\jname{Dynamics of Solar and Stellar \\ Convection Zones and Atmospheres}
\editors{A.C. Editor, B.D. Editor \& C.E. Editor, eds.}
\begin{document}

\maketitle

\begin{abstract}
To create early warning capabilities for upcoming Space Weather disturbances, we have selected a dataset of 61 emerging active regions, which allows us to identify characteristic features in the evolution of acoustic power density to predict continuum intensity emergence. For our study, we have utilized Doppler shift and continuum intensity observations from the Helioseismic and Magnetic Imager (HMI) onboard the Solar Dynamics Observatory (SDO). The local tracking of $30.66 \times 30.66$-degree patches in the vicinity of active regions allowed us to trace the evolution of active regions starting from the pre-emergence state. We have developed a machine learning model to capture the acoustic power flux density variations associated with upcoming magnetic flux emergence. The trained Long Short-Term Memory (LSTM) model is able to predict 5 hours ahead whether, in a given area of the solar surface, continuum intensity values will decrease. The performed study allows us to investigate the potential of the machine learning approach to predict the emergence of active regions using acoustic power maps as input.
\end{abstract}
\firstsection 
\section{Introduction}

Building early forecasts of the manifestation of solar activity is crucial to mitigate the impact of Space weather disturbances. Since most research efforts focus on predicting eruptive activities or their consequences, we consider investigating the potential of a machine learning approach to predict the appearance of its source. Capturing signatures of the subsurface activities will allow researchers to anticipate surface phenomena like magnetic flux emergence followed by the formation of eruptive active regions. The initial attempt to observe an emerging active region in the solar interior was made by \cite{kosovichev2001time}, who used time-distance helioseismology to image sound-speed variations prior to the appearance of the active region on the surface. The results showed the emerging magnetic flux is fragmented and travels very quickly through the top 18 Mm layer of the convection zone, with a speed of about 1 km/s. The feasibility of using variations in acoustic power as a precursor to detect signatures of emerging magnetic flux was initially explored by \cite{hartlep2011signatures}. Using numerical simulations of acoustic wave propagation within the solar interior, applied to AR10488, they demonstrated that subsurface structures could influence the acoustic power at the overlying photosphere. These findings suggested that changes in acoustic power could be indicative of active regions before they become visible on the surface. 

Subsequent studies further explored the potential of helioseismology in detecting pre-emergence signatures of solar active regions. For instance, \cite{leka2012helioseismology} analyzed over 100 instances of emerging and non-emerging solar regions, using data from the Global Oscillations Network Group \citep[GONG]{harvey1996global} and SOHO/MDI \citep{scherrer1995solar}. \cite{birch2012helioseismology} used helioseismic holography on these active region samples and identified statistically significant differences in subsurface flows and wave speeds preceding active region formation. Their findings effectively ruled out extensive flows above 15~m/s in the top 20 Mm of the photosphere prior to emergence, thereby setting constraints on theoretical models of active region development. \cite{barnes2014helioseismology} performed a statistical analysis of these ARs' subsurface properties, highlighting the surface magnetic field as a key differentiator between the two populations of regions on the Sun (pre-emergence and without emergence). The research further revealed helioseismic precursors to active region emergence, detectable at least a day prior. These works showed that the subsurface dynamics play an important role in solar active region formation and also acknowledged the limitations posed by the small sample sizes in definitively determining these subsurface properties. 

The possibility of using synoptic imaging of subsurface magnetic activity in order to predict large sunspot regions before they become visible was demonstrated by \cite{ilonidis2011detection}, who detected strong acoustic travel-time anomalies of an order of 12 to 16 seconds as deep as 65~Mm. These anomalies, originating from deeper layers, rise to shallower regions at velocities up to 1~km/s, suggesting their association with acoustic power variations rather than just subsurface flows or wave-speed perturbations \citep{ilonidis2013helioseismic}. 
Similarly, \cite{stefan2023exploring} investigated the deviations in the mean phase travel time of acoustic waves before the emergence of 46 large active regions, showing the relationship between subsurface acoustic signals and surface magnetic flux. Despite some limitations in the strength of correlations and the variability in lag times, their findings underscore a discernible connection between the mean phase travel times and the emergence of active regions. 

In a similar vein, \cite{attie2018precursors} observed disruption of the moat flow several hours before the onset of strong flux emergence near active region AR12673. This disruption occurred at the same location where magnetic flux would later emerge. Horizontal divergent flows at the solar surface have also been identified as potential precursors to flux emergence. Using SDO/HMI data, \cite{rees2022preemergence} showed that about 37.5\% of active regions display divergent flows divergent flows 58 minutes before their emergence. Converging flows with a speed of 20-30~m/s one day before the emergence was also observed by \cite{gottschling2021evolution}.


Strengthening of the solar f-mode prior to active region emergence was observed by \cite{waidele2023strengthening}, who utilized the Fourier-Hankel analysis to examine six active regions. This noticeable enhancement in the f-mode power, one to three days before the emergence of these regions, corroborates earlier findings by \cite{singh2016high}. However, this enhancement varied significantly across different regions. Importantly, the analysis by \cite{waidele2023strengthening} did not find any directional dependence (in terms of inward and outward moving waves) associated with this f-mode power strengthening. This non-directional nature of the phenomenon and the lack of a strong correlation with surface magnetic flux evolution highlights the complexity of subsurface processes preceding active region emergence. Further analysis by \cite{korpi2022solar}, who applied a new calibration technique, did not reveal a significant enhancement of the f-mode prior to AR emergence.

In this paper, we explore the possibility of detecting emerging active regions before they become visible on the solar surface using a machine learning method trained on observations from the HMI instrument onboard SDO. 

\section{Data Preparation}

\begin{figure}[t]
\centering
\includegraphics[width=\linewidth]{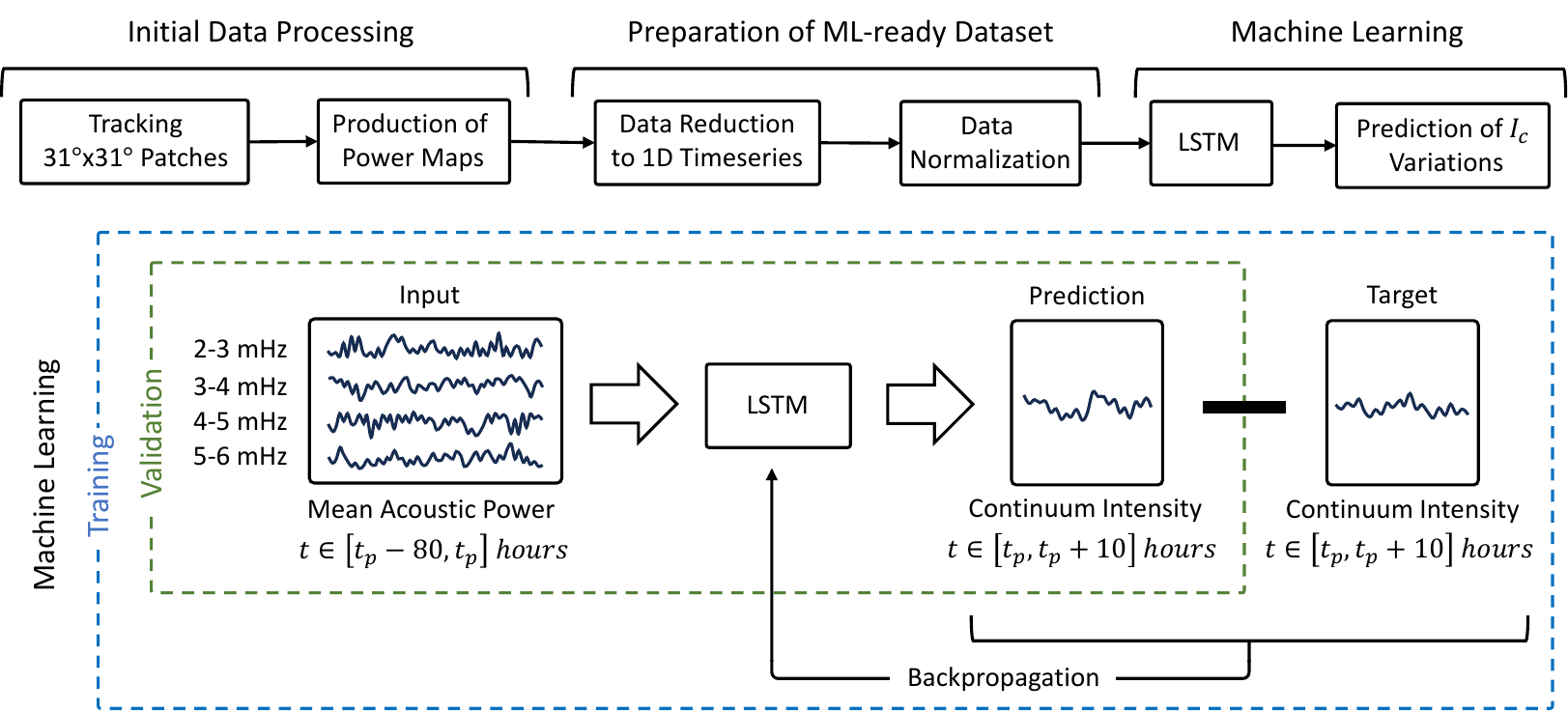}
\caption{Scheme of the workflow (top panel) and the machine learning operations (bottom panel) to predict the emergence of active regions. Denoted at $t_p$ is the time of prediction in each time step of the LSTM rolling window approach.}
\label{fig:pipeline}
\end{figure}

To investigate capabilities to predict the appearance of active regions on the solar surface, we analyze the evolution of 61 active regions (ARs) using data from the Helioseismic and Magnetic Imager \citep[HMI]{scherrer2012helioseismic} onboard the Solar Dynamics Observatory (SDO). Each selected active region emerged on the solar surface within 30 degrees longitude from the central meridian between 1 March 2010 and 1 June 2023, lived for longer than 4 days, and reached a total area of 200 millionth of the solar hemisphere. The longitude range was set to minimize significant distortion due to the projection and the center-to-limb effects. Our analysis includes three main steps as presented in Figure~\ref{fig:pipeline}: a) initial data processing, b) preparation of the ML-ready input data and continuum intensity time series to verify results, and c) development and training of an ML-driven model to predict the appearance of active regions using the Long Short-Term Memory (LSTM) method.

To perform data analysis, we tracked the corresponding $30.66\times30.66$ degrees area with local rotation speed (Figure~\ref{fig:pipeline}). These patches represent a $512\times512$-pixel square centered around the target AR. The temporal tracking of ARs is executed through a division into overlapping 8-hour time series. Each of them is comprised of 640 frames, with cadence of 45 seconds. This approach is applied to all HMI data products, including Dopplergrams, line-of-sight magnetograms, and continuum intensity (${\rm I_c}$) maps. The Dopplergrams time series have been used to generate acoustic power maps for four frequency ranges: 2–3, 3–4, 4–5, and 5–6 mHz for the entire 10-day period of the passage of the ARs areas before and after their emergence.

\begin{figure}[h]
\centering
\includegraphics[width=\linewidth]{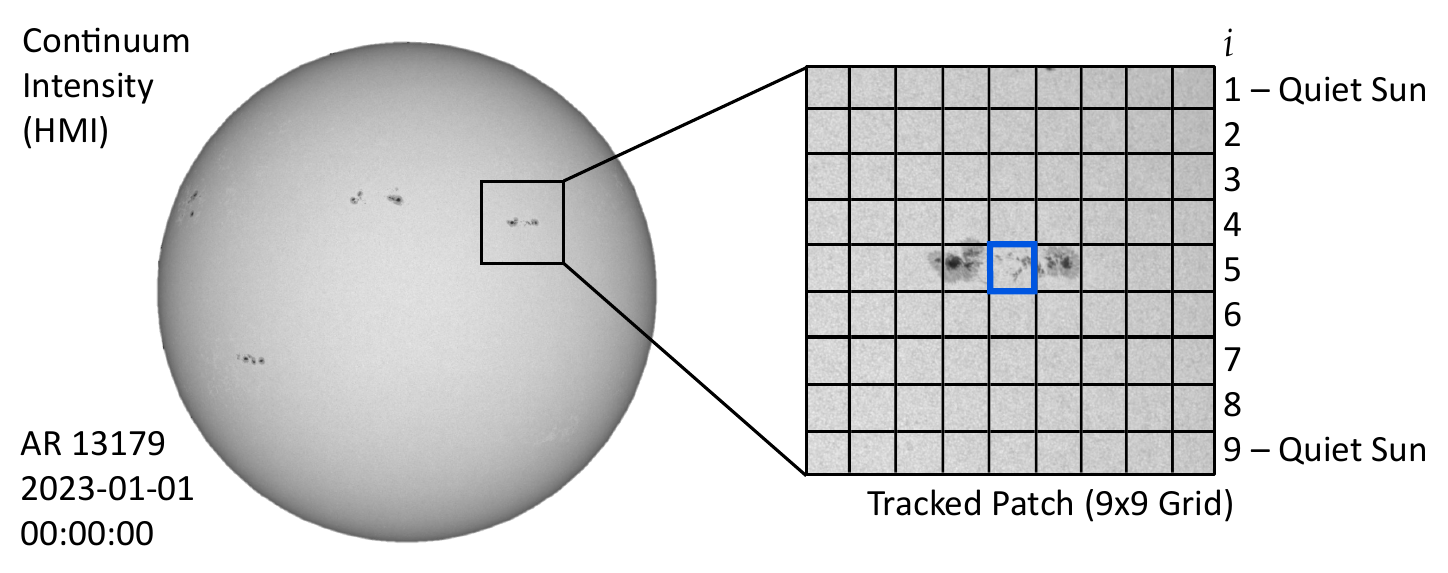}
\caption{The tracked $30.66\times30.66$ degree patch for AR13179 (left) within the HMI continuum intensity full disc and the 9 by 9 grid it was split in (right). The colored patch corresponds to the variations of the acoustic power and continuum intensity shown in the same color in Figure~\ref{fig:acoustic_power_timelines}.}
\label{fig:normalization}
\end{figure}

The resulting power maps, magnetograms, and intensity maps were split into a grid of 9 by 9 tiles (Figure~\ref{fig:normalization}). The grid size ($N$) is adjustable and can be used for further exploration to increase the granularity of data, offer improved prediction accuracy, and provide details of the emergence scenario. This splitting process enabled tracing the evolution of the local area and, by calculating the mean value of each tile, it served as a data reduction step (third step, Figure~\ref{fig:pipeline}), which created structured timelines conducive to machine learning analysis. Because the tracking area was selected with the AR at the center (determined from the NOAA AR database), the grid size was chosen for convenience of analysis so that the central tile typically corresponds to the center of the developed AR. Thus, dividing a large area into small tiles allows us to investigate the possibility of predicting the time and location of an active region formation.

\begin{figure}[h]
\centering
\includegraphics[width=\linewidth]{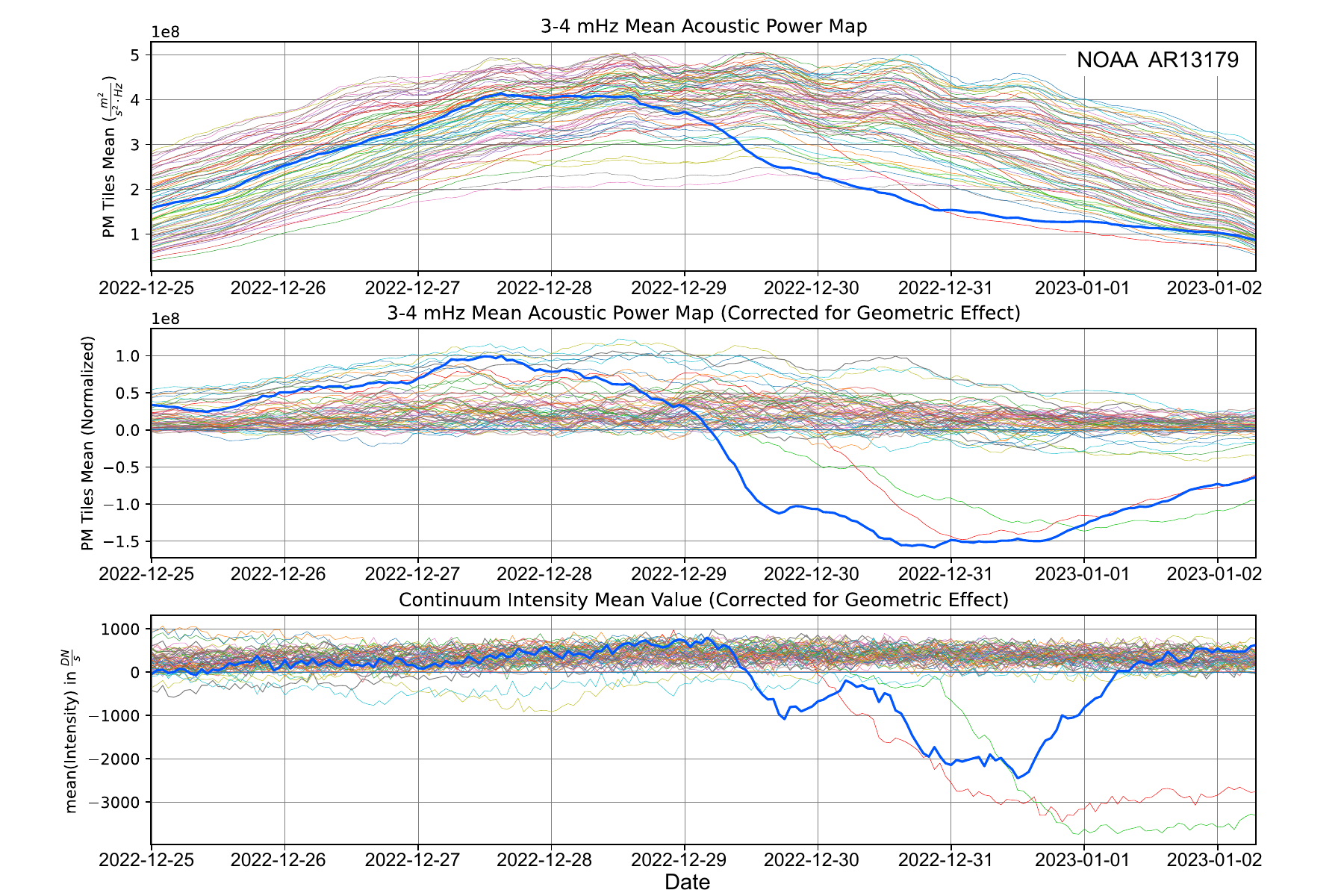}
\caption{Time evolution of the mean acoustic power for 63 tiles obtained from the $30.66\times30.66$-degrees area for the 3-4~mHz frequency range (top panel). The middle and bottom panels show the evolution of the normalized acoustic power and the continuum intensity corrected for the geometric effect. The blue curve shows the emergence and evolution of AR12085 in the same color-coded middle tile in Figure~\ref{fig:normalization}.}
\label{fig:acoustic_power_timelines}
\end{figure}

In the current 9 by 9 configuration, each tile represents a 57 by 57~pixel area, corresponding to 3.36~degrees of heliographic longitude and latitude. After taking the mean over each tile for acoustic power and continuum intensity maps, we obtained an ensemble of time series (or timelines) that describes the evolution of the area within the tile. Similarly, we characterize the local evolution for unsigned magnetic flux. We excluded 20 active regions out of 61 from ML analysis because of data gaps and quality issues that require additional processing and investigation.
The resulting timelines in Figure~\ref{fig:acoustic_power_timelines}, provide the structured time series data necessary for ML-driven forecasts. 

To remove the geometric effect, which manifests as systematic variations in the resulting timelines (Figure~\ref{fig:acoustic_power_timelines}, top panel), we normalize the data using the most distant tiles at the same longitude (top and bottom tiles marked as ``Quite Sun" in Figure~\ref{fig:normalization}). 
Because in our dataset these tiles typically represent the quiet-Sun regions, we use them as a baseline to normalize the tiles in the intermediate latitudes relative to zero. These `reference' tiles are excluded from the ML analysis, bringing the number of tiles (and therefore timelines) for each active region from 81 down to 63. Normalization of the timelines has been performed along the longitudinal direction (Figure~\ref{fig:normalization}), using the following expression

\begin{equation}
P_i = P_i - \left( \frac{i}{N} \cdot P_1 + (1 - \frac{i}{N}) \cdot P_N \right),\nonumber
\label{eq:Norm}
\end{equation}
where $P_i$ is a row number for $i=1, 2, ..., N$. In our case, $N=9$.

The subtraction of the temporal variations due to passing the tracking area across the solar disk and the normalization based on the quiet-Sun region values allows us to amplify the variations related to the magnetic activity (Figure~\ref{fig:acoustic_power_timelines}, middle and bottom panels). Because the reference tiles, which are assumed to be `Quiet Sun' regions, may exhibit some activity that can contribute to these deviations. Significant variations of the mean power may indicate changes in the properties of the acoustic waves and turbulence due to the coming large-scale magnetic flux from solar interiors. Before they are used to train our ML model, the acoustic power maps and intensity time series are further normalized to a range from 0 to 1 to adapt to our LSTM model. In the following section, we discuss the LSTM machine learning method to investigate acoustic power suppression as a precursor of emerging magnetic flux and the formation of an active region.

\section{Long Short-Term Memory Modeling}

Because of the complex links between turbulent motions on the solar photosphere, small-scale dynamo, coming flows, and new magnetic fields from the solar interior, capturing structural and dynamical changes due to the emergence of an active region is not a trivial problem. To address this, we use machine learning to advance the cross-analysis of the local time evolution of the acoustic power for four frequency ranges (2–3, 3–4, 4–5, and 5–6 mHz) and predict a decrease in the mean continuum intensity associated with the appearance of an active region on the solar surface (Figure~\ref{fig:pipeline}, bottom panel). An example of the data set for the 3-4 mHz frequency range is shown in Figure~\ref{fig:acoustic_power_timelines}, middle panel. 

The strategy to utilize the variations of the oscillatory properties of the solar convection to reconstruct and predict the continuum intensity is based on a general approach to detect active regions at the photosphere when the AR appearance always accompanies the intensity decrease. Because including information about a local evolution of magnetic flux will enhance the predictive capabilities, we will use magnetograms to improve the predictive capabilities of the current ML model. However, using the unsigned magnetic flux as a criterion to define the emergence of an active region makes the results more dependent on a selection of tile size and more sensitive to the enhancement of the magnetic energy on smaller scales, which are out of the scope of this study.

In this work, we employ Long Short-Term Memory \citep{hochreiter1997long} networks, a type of recurrent neural network (RNN) that is particularly adept at handling sequential data \citep{hua2019deep,tealab2018time,sak2014long} and aims to resolve the vanishing gradient problem \citep{basodi2020gradient}. LSTM networks are designed to overcome the limitations of traditional RNNs, particularly in learning long-term dependencies. This is achieved through their unique architecture, which includes memory cells and multiple gates (input, output, and forget gates). These components allow LSTM to regulate the flow of information, effectively remembering important inputs over long sequences while discarding irrelevant data. This capability makes LSTM beneficial for modeling and predicting the local dynamics at the solar surface and, in particular, forecasting the formation and evolution of ARs based on the intricate patterns observed in our acoustic power density and intensity timelines.

The LSTM method employs a sliding window technique to analyze local temporal variations from the 1D time series of the acoustic power to estimate current and future continuum intensity variations. To perform analysis, the 80-hour time window consequently shifted by 1~hour for different prediction times $t_p$ as seen in Figure~\ref{fig:pipeline}. During the processing of each time window, the LSTM model learned how to predict the mean intensity variations on short-time scales (next 5 hours) using the recent history of acoustic power evolution across different frequencies. During validation (Figure~\ref{fig:pipeline}), the same process is employed, with the only difference being that this time, the resulting prediction outcome is not compared to the true intensity values, and therefore, this information is not backpropagated to the LSTM and its weights (trainable parameters) are not altered. This analysis is applied to a time series of four frequency ranges and spans across 63 tiles, regardless of their activity level for 40 ARs.

In the LSTM model, we use a hidden layer size of 64, which refers to the number of hidden units within each layer. Every hidden unit (or cell) can be considered as a memory cell that processes input data, retains important information, and passes this information to the next time step or the next layer. Thus, it influences the network's capacity to learn complex patterns. In our configuration, the LSTM comprises three layers, keeping a balance between model complexity and computational efficiency. Our model's learning rate is set at 0.001, while the number of epochs (number of times the training process is repeated) is set to 500, allowing for a slow but steady convergence to the optimal trainable parameter space using the gradient descent approach, with Mean Squared Error (MSE) loss function guiding the model's accuracy improvement. The MSE is a widely used cost function that calculates the average squared difference between the predicted and actual values, providing a measure of the model's prediction accuracy. This relatively low number of epochs also determines the duration of the model training, allowing the LSTM to iteratively refine its predictions in less than 12 hours when run on an NVIDIA A100 GPU node.

\begin{figure}[b]
    \centering
    \includegraphics[width=\linewidth]{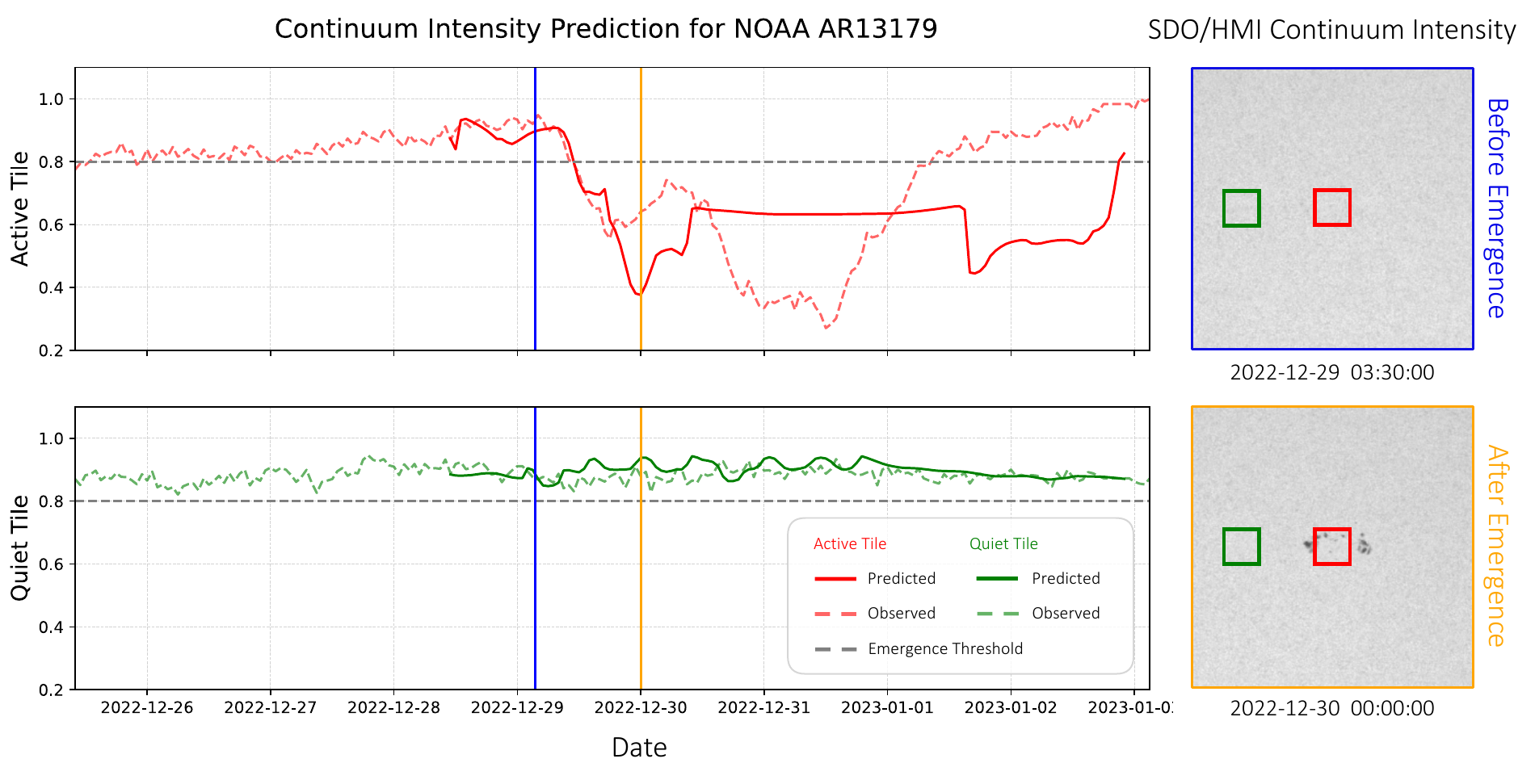}
    \caption{Continuum intensity prediction for AR13179 by the trained LSTM model. The left panels show the predicted (solid curves) and actual (dashed) continuum intensity variations in tiles marked on the right continuum intensity images. The vertical lines indicate the moment half an hour before emergence (blue) and a moment after emergence (orange). The right panels show the whole 30.66 by 30.66~degrees continuum intensity areas corresponding to these moments before (top) and after emergence (bottom image). In the snapshots on the right, enclosed in a red square is a tile that exhibits the formation of an active region (active tile, upper plot) and corresponds to the blue middle tile in Figure~\ref{fig:normalization}. Enclosed in the green square is a quiet-Sun region (quiet tile, bottom plot). The grey horizontal dashed line indicates the threshold that intensity has to exceed in order to have emergence in this study.}
    \label{fig:lstm_results}
\end{figure}

\section{Prediction of the Continuum Intensity Variations}

In this work, we assume that the emergence of the AR starts when the following criteria for the mean normalized intensity are satisfied: 1) intensity decreases below the threshold of 0.8; 2) the time derivative is -0.2 or less; and 3) the previous criteria are satisfied for at least 12 hours. We illustrate the LSTM machine learning model performance for two examples to predict the evolution of the continuum intensity for a tile which manifested the formation of AR13179, a so-called `active' tile (red rectangle in Figure~\ref{fig:lstm_results}) and quiet-Sun region tile referred to as `quiet' (green). According to NOAA, the emergence of AR13179 was observed on 29 December 2022, 04:00:00 TAI. To predict intensity for a particular time, the intensity timeline of at least 85~hours is required (80~hours is the defined LSTM input plus 5~hours not seen by the LSTM for the time window). Therefore, in our study, we begin predicting intensity starting at 10:00:00 on 28 December 2022. 

Prediction of the continuum intensity for 5 hours ahead (solid curve, Figure~\ref{fig:lstm_results}, bottom plot) that does not exhibit magnetic activity (marked as a green tile) shows a good agreement with the observations (dashed curve). For the active tile, the LSTM model (Figure~\ref{fig:lstm_results}, upper plot) can reproduce well a general trend of intensity decrease, especially at the beginning of the emergence. It is important to note that newly formed and evolving magnetic structures often do not stay in the same place and can migrate from one tile to another. Also, during the emergence and formation of an active region, the local area experiences structural and dynamic changes in flows and magnetic fields, expands to a larger area, and occupies several tiles. The LSTM model is not trained on topological changes of magnetic fields and associated with it the oscillatory properties of the convection. Therefore, discrepancies between the predicted and observed intensity during AR emergence may increase with time (Figure~\ref{fig:lstm_results}, upper plot). Nevertheless, the success of capturing variations of the continuum intensity during the beginning of the flux emergence creates confidence in the ability to apply the LSTM model to predict the appearance of active regions on the solar surface before they become visible.

\section{Discussion and Conclusions}

To investigate the applicability of the Long Short-Memory (LSTM) algorithm to predict the formation of an active region before it becomes visible on the solar surface, we analyzed 40 active regions observed by SDO/HMI. The formation of active regions can easily be observed in the continuum intensity. Previously founded correlations between changes in the acoustic power and upcoming magnetic activity on the solar surface suggest that this coupling can serve as a reliable predictor. However, the non-linear nature of these correlations requires advanced statistical analysis. Therefore, we developed an LSTM model to predict intensity variations based on local changes in the mean acoustic power. The training was performed for 40 active regions, where time series of the mean continuum intensity and acoustic power for four frequency ranges were utilized. We analyze the time series over 63 tiles for each active region, covering quiet-Sun regions and upcoming emergence areas. The LSTM model was constructed with 64 hidden layers and used a learning rate of 0.001 for 500 epochs.

Our validation results demonstrate the capability of an ML approach and, in particular, the LSTM method to predict emergence at least 5 hours in advance for the active areas of AR13179, which is not used during training and, therefore, we use it for validation. Predicting the continuum intensity variations before and at the beginning of active region formation using changes in the acoustic power as input shows remarkable results even if information about magnetic fields is not available to the model. The large discrepancies between the prediction and observations during the late stage of active region formation suggest that the training strategy requires additional investigation.
Thus, the model's capability to accurately forecast significant changes in solar activity underscores its potential as a powerful tool for space weather forecasting. Future efforts will be concentrated on increasing the LSTM model predictive capabilities by including training on magnetic field observations, improving data calibration, and optimizing the model parameters.

{\bf Acknowledgements}\\
We want to thank the NAS Visualization Team (Nina McCurdy, Timothy Sandstrom, Christopher Henze) for their help with this project's visualizations. This work is supported by the NASA AI/ML HECC Expansion Program, NASA Heliophysics Supporting Research Program, and the NASA grants 80NSSC19K0630, 80NSSC19K0268, 80NSSC20K1870, and 80NSSC22M0162.


\begin{thebibliography}{}

\bibitem[Atti{\'e} et~al., 2018]{attie2018precursors}
Atti{\'e}, R., Kirk, M.~S., Thompson, B.~J., Muglach, K., \& Norton, A.~A.
  2018, Precursors of magnetic flux emergence in the moat flows of active
  region ar12673.
\newblock {\em Space Weather}, 16(8), 1143--1155.

\bibitem[Barnes et~al., 2014]{barnes2014helioseismology}
Barnes, G., Birch, A., Leka, K., \& Braun, D. 2014, Helioseismology of
  pre-emerging active regions. iii. statistical analysis.
\newblock {\em The Astrophysical Journal}, 786(1), 19.

\bibitem[Basodi et~al., 2020]{basodi2020gradient}
Basodi, S., Ji, C., Zhang, H., \& Pan, Y. 2020, Gradient amplification: An
  efficient way to train deep neural networks.
\newblock {\em Big Data Mining and Analytics}, 3(3), 196--207.

\bibitem[Birch et~al., 2012]{birch2012helioseismology}
Birch, A., Braun, D., Leka, K., Barnes, G., \& Javornik, B. 2012,
  Helioseismology of pre-emerging active regions. ii. average emergence
  properties.
\newblock {\em The Astrophysical Journal}, 762(2), 131.

\bibitem[Gottschling et~al., 2021]{gottschling2021evolution}
Gottschling, N., Schunker, H., Birch, A., L{\"o}ptien, B., \& Gizon, L. 2021,
  Evolution of solar surface inflows around emerging active regions.
\newblock {\em Astronomy \& Astrophysics}, 652, A148.

\bibitem[Hartlep et~al., 2011]{hartlep2011signatures}
Hartlep, T., Kosovichev, A.~G., Zhao, J., \& Mansour, N.~N. 2011, Signatures of
  emerging subsurface structures in acoustic power maps of the sun.
\newblock {\em Solar Physics}, 268, 321--327.

\bibitem[Harvey et~al., 1996]{harvey1996global}
Harvey, J., Hill, F., Hubbard, R., Kennedy, J., Leibacher, J., Pintar, J.,
  Gilman, P., Noyes, R., Title, A., Toomre, J., {\it et~al.} 1996, The global
  oscillation network group (gong) project.
\newblock {\em Science}, 272(5266), 1284--1286.

\bibitem[Hochreiter and Schmidhuber, 1997]{hochreiter1997long}
Hochreiter, S. \& Schmidhuber, J. 1997, Long short-term memory.
\newblock {\em Neural computation}, 9(8), 1735--1780.

\bibitem[Hua et~al., 2019]{hua2019deep}
Hua, Y., Zhao, Z., Li, R., Chen, X., Liu, Z., \& Zhang, H. 2019, Deep learning
  with long short-term memory for time series prediction.
\newblock {\em IEEE Communications Magazine}, 57(6), 114--119.

\bibitem[Ilonidis et~al., 2013]{ilonidis2013helioseismic}
Ilonidis, S., Zhao, J., \& Hartlep, T. 2013, Helioseismic investigation of
  emerging magnetic flux in the solar convection zone.
\newblock {\em The Astrophysical Journal}, 777(2), 138.

\bibitem[Ilonidis et~al., 2011]{ilonidis2011detection}
Ilonidis, S., Zhao, J., \& Kosovichev, A. 2011, Detection of emerging sunspot
  regions in the solar interior.
\newblock {\em Science}, 333(6045), 993--996.

\bibitem[Korpi-Lagg et~al., 2022]{korpi2022solar}
Korpi-Lagg, M.~J., Korpi-Lagg, A., Olspert, N., \& Truong, H.-L. 2022,
  Solar-cycle variation of quiet-sun magnetism and surface gravity oscillation
  mode.
\newblock {\em Astronomy \& Astrophysics}, 665, A141.

\bibitem[Kosovichev et~al., 2001]{kosovichev2001time}
Kosovichev, A., Duvall, T., \& Scherrer, P. 2001, Time-distance inversion
  methods and results (invited review).
\newblock {\em Helioseismic Diagnostics of Solar Convection and Activity},,
  159--176.

\bibitem[Leka et~al., 2012]{leka2012helioseismology}
Leka, K., Barnes, G., Birch, A., Gonzalez-Hernandez, I., Dunn, T., Javornik,
  B., \& Braun, D. 2012, Helioseismology of pre-emerging active regions. i.
  overview, data, and target selection criteria.
\newblock {\em The Astrophysical Journal}, 762(2), 130.

\bibitem[Rees-Crockford et~al., 2022]{rees2022preemergence}
Rees-Crockford, T., Nelson, C., \& Mathioudakis, M. 2022, Preemergence
  signatures of horizontal divergent flows in solar active regions.
\newblock {\em The Astrophysical Journal}, 940(2), 109.

\bibitem[Sak et~al., 2014]{sak2014long}
Sak, H., Senior, A.~W., \& Beaufays, F. 2014, Long short-term memory recurrent
  neural network architectures for large scale acoustic modeling.

\bibitem[Scherrer et~al., 1995]{scherrer1995solar}
Scherrer, P.~H., Bogart, R.~S., Bush, R., Hoeksema, J.-a., Kosovichev, A.,
  Schou, J., Rosenberg, W., Springer, L., Tarbell, T., Title, A., {\it et~al.}
  1995, The solar oscillations investigation—michelson doppler imager.
\newblock {\em The soho mission},, 129--188.

\bibitem[Scherrer et~al., 2012]{scherrer2012helioseismic}
Scherrer, P.~H., Schou, J., Bush, R., Kosovichev, A., Bogart, R., Hoeksema, J.,
  Liu, Y., Duvall, T., Zhao, J., Title, A., {\it et~al.} 2012, The helioseismic
  and magnetic imager (hmi) investigation for the solar dynamics observatory
  (sdo).
\newblock {\em Solar Physics}, 275, 207--227.

\bibitem[Singh et~al., 2016]{singh2016high}
Singh, N.~K., Raichur, H., \& Brandenburg, A. 2016, High-wavenumber solar
  f-mode strengthening prior to active region formation.
\newblock {\em The Astrophysical Journal}, 832(2), 120.

\bibitem[Stefan and Kosovichev, 2023]{stefan2023exploring}
Stefan, J.~T. \& Kosovichev, A.~G. 2023, Exploring the connection between
  helioseismic travel time anomalies and the emergence of large active regions
  during solar cycle 24.
\newblock {\em The Astrophysical Journal}, 948(1), 1.

\bibitem[Tealab, 2018]{tealab2018time}
Tealab, A. 2018, Time series forecasting using artificial neural networks
  methodologies: A systematic review.
\newblock {\em Future Computing and Informatics Journal}, 3(2), 334--340.

\bibitem[Waidele et~al., 2023]{waidele2023strengthening}
Waidele, M., Roth, M., Singh, N., \& K{\"a}pyl{\"a}, P. 2023, On strengthening
  of the solar f-mode prior to active region emergence using the fourier-hankel
  analysis.
\newblock {\em Solar Physics}, 298(2), 30.
\end{thebibliography}

\end{document}